\documentstyle[12pt,epsfig]{article}
\begin{document}
\newcommand{\beq}{\begin{equation}}
\newcommand{\eeq}{\end{equation}}
\newcommand{\barr}{\begin{eqnarray}}
\newcommand{\earr}{\end{eqnarray}}

\newcommand{\andy}[1]{ }

\def\txt{\textstyle}
\def\eqn#1{Eq.\ (\ref{eq:#1})} \def\pp{{\vec p}}
\def\ask{\marginpar{?? ask:  \hfill}}
\def\fin{\marginpar{fill in ... \hfill}}
\def\note{\marginpar{note \hfill}}
\def\check{\marginpar{check \hfill}}
\def\discuss{\marginpar{discuss \hfill}}
\def\hh{\widehat}
\def\tt{\widetilde}
\def\cH{{\cal H}}
\def\cT{{\cal T}}
\def\cR{{\cal R}}
\def\cC{{\cal C}}
\def\cZ{{\cal Z}}
\def\cL{{\cal L}}
\def\txt{\textstyle}
\def\bmp{\mbox{\boldmath $p$}}
\def\bmk{\mbox{\boldmath $k$}}
\def\bmksub{\mbox{\boldmath\scriptsize $k$}}
\def\bmp{\mbox{\boldmath $p$}}
\def\bmr{\mbox{\boldmath $r$}}
\def\bmA{\mbox{\boldmath $A$}}
\def\bmv{\mbox{\boldmath $v$}}
\def\bmg{\mbox{\boldmath $g$}}
\def\bmepsilon{\mbox{\boldmath $\epsilon$}}
\def\bmhA{\mbox{\boldmath $\hat{A}$}}
\def\bmhp{\mbox{\boldmath $\hat{p}$}}
\def\bmhv{\mbox{\boldmath $\hat{v}$}}
\def\pade#1#2{{frac{\partial#1}{\partial#2}}}
\newcommand{\mean}[1]{\langle #1 \rangle}



\begin{titlepage}
\begin{flushright}
\today \\
BA-TH/99-344\\
\end{flushright}
\vspace{.5cm}
\begin{center}
{\LARGE Quantum chaos induced by measurements\footnote{To appear in
the Proceedings of the Symposium on ``Mathematical Aspects of
Quantum Information and Quantum Chaos," Kyoto, Japan, 1999.}

}

\quad

{\large P. FACCHI,$^{(a,b)}$ S. PASCAZIO$^{(a,b)}$ and A.
SCARDICCHIO$^{(a)}$\\
           \quad    \\

        $^{(a)}$Dipartimento di Fisica, Universit\`a di Bari
 I-70126  Bari, Italy

       $^{(b)}$Istituto Nazionale di Fisica Nucleare, Sezione di Bari \\
 I-70126  Bari, Italy \\

}

\vspace*{.5cm} PACS: 05.45.Mt; 05.45.-a; 05.40.-a; 03.65.Bz
\vspace*{.5cm}

{\small\bf Abstract}\\ \end{center}

{\small

We study the dynamics of a ``kicked" quantum system undergoing
repeated measurements of momentum. A diffusive behavior is obtained
for a large class of Hamiltonians, even when the dynamics of the
classical counterpart is not chaotic. These results can be
interpreted in classical terms by making use of a ``randomized"
classical map. We compute the transition probability for the
action variable and consider the semiclassical limit.

}

\end{titlepage}

\newpage

\setcounter{equation}{0}
\section{Introduction }
 \label{sec-introd}
 \andy{intro}

The classical and quantum dynamics of bound Hamiltonian systems
under the action of periodic ``kicks" are in general very
different. Classical systems can follow very complicated
trajectories in phase space, while the evolution of the wave
function in the quantum case is more regular. In the classical
case, in those regions of the phase space that are stochastic, the
evolution of the system can be well described in terms of the
action variable alone and one of the most distinctive features of
an underlying chaotic behavior is just the diffusion of the action
variable in phase space. On the other hand, in the quantum case,
such a diffusion is always suppressed after a sufficiently long
time \cite{Chirikov,Berry}. This phenomenon, known as the quantum
mechanical suppression of classical chaos, can be framed in a
proper context in terms of the semiclassical approximation $\hbar
\to 0$ \cite{CCSG,Tabor}.

The ``kicked" rotator is a pendulum that evolves under the action
of a gravitational field that is ``switched on" at periodic time
intervals. It is a very useful system, able to elucidate many
different features between the classical and the quantum case. By
studying this model, Kaulakis and Gontis \cite{KaulGontis} showed
that a diffusive behavior of the action variable takes place even
in the quantum case, if a quantum measurement is performed after
every kick. This interesting observation was investigated in some
detail in a recent paper \cite{FPS}, where it was proven that
quantum mechanical measurements of the action variable provoke
diffusion in a very large class of ``kicked" systems, even when the
corresponding classical dynamics is regular. In this paper we shall
first briefly review some of our general results and then
corroborate our findings by concentrating our attention on the
particular case of the kicked rotator.

\setcounter{equation}{0}
\section{Kicks interspersed with quantum measurements }
 \label{sec-kickmeas}
 \andy{kickmeas}

We consider the Hamiltonian
\andy{eq:hamilt}
\beq
H=H_0(p) + \lambda V(x) \delta_T(t),
\label{eq:hamilt}
\eeq
where $p$ and $x \in [-\pi,\pi]$ are the action and angle variable,
respectively, and
\andy{eq:deltat}
\beq
\delta_T(t)=\sum_{k=-\infty}^{\infty} \delta (t-kT),
\label{eq:deltat}
\eeq
$T$ being the period of the perturbation. We impose periodic
boundary conditions on the interaction $V(x)$. This Hamiltonian
gives rise to the so-called radial twisting map, that describes the
local behavior of a perturbed integrable map near resonance
\cite{Lichten}. The free Hamiltonian $H_0$ has a discrete spectrum
and a countable complete set of eigenstates $\{|m\rangle\}$:
\andy{eigenfun}
\beq
\langle x|m\rangle=\frac{1}{\sqrt{2\pi}} \exp\left(imx\right),
\qquad m=0,\pm1,\pm2,\ldots\ .
\label{eq:eigenfun}
\eeq
We shall consider the evolution engendered by the Hamiltonian
(\ref{eq:hamilt}) interspersed with quantum measurements, in the
following sense: the system evolves under the action of the free
Hamiltonian for $(N-1)T+\tau<t< NT \;$ ($0<\tau<T$), undergoes a
``kick" at $t=NT$, evolves again freely and then undergoes a
``measurement" of $p$ at $t=NT+\tau$. The evolution of the system
is best described in terms of the density matrix: between
successive measurements one has
\andy{eq:propag1,2}
\barr
\rho_{NT+\tau}&=&U_{\rm free}(\tau) U_{\rm kick} U_{\rm free}(T-\tau)
\rho_{(N-1)T+\tau}U_{\rm free}^\dagger(T-\tau)U_{\rm kick}^\dagger
U_{\rm free}^\dagger(\tau),
 \label{eq:propag1} \\
& & U_{\rm kick} = \exp\left(-i \lambda V/ \hbar \right), \quad
U_{\rm free} (t) = \exp\left(-i H_0 t/ \hbar\right).
\label{eq:propag2}
\earr
At each measurement, the wave function is ``projected" onto the
$n$th eigenstate of $p$ with probability $P_n(NT+\tau)={\rm
Tr}(|n\rangle\langle n|\rho_{NT+\tau})$ and the off-diagonal terms
of the density matrix disappear. The occupation probabilities
$P_n(t)$ change discontinuously at times $NT$ and their evolution
is governed by the master equation
\andy{eq:master}
\beq
P_n(N)=\sum_m W_{nm}P_m(N-1),
\label{eq:master}
\eeq
where
\andy{eq:transprob}
\beq
W_{nm} \equiv
|\langle n|U_{\rm free}(\tau)U_{\rm kick}U_{\rm free}(T-\tau)|m \rangle|^2=|\langle n|U_{\rm
kick}|m \rangle|^2
\label{eq:transprob}
\eeq
are the transition probabilities and we defined, with a little
abuse of notation,
\andy{defP}
\beq
P_n(N) \equiv P_n(NT+\tau).
\label{defP}
\eeq
The map (\ref{eq:master}) depends on $\lambda, V, H_0$ in a
complicated way. However, interestingly, very general conclusions
can be drawn about the average value of a generic regular function
of momentum $g(p)$ \cite{FPS}. Let
\andy{eq:energy}
\beq
\mean{g(p)}_t \equiv {\rm Tr}(g(p)\rho(t))=
\sum_n g(p_n) P_n(t),
\label{eq:energy}
\eeq
where $p |n\rangle = p_n |n \rangle$ $(p_n=n\hbar)$, and consider
the average value of $g$ after $N$ kicks
\andy{eq:energy1}
\beq
\mean{g(p)}_{N} \equiv
\mean{g(p)}_{NT+\tau}
=\sum_n g(p_n) P_n(N)=\sum_{n,m} g(p_n) W_{nm}P_m(N-1).
\label{eq:energy1}
\eeq
Substituting $W_{nm}$ from (\ref{eq:transprob}) one obtains
\andy{eq:energy2}
\barr
\mean{g(p)}_{N} &=&
\sum_{n,m} g(p_n) \langle m|U_{\rm kick}^\dagger|n\rangle
\langle n|U_{\rm kick}|m\rangle P_m(N-1)\nonumber \\
&=& \sum_m \langle m|U_{\rm kick}^\dagger g(p) U_{\rm kick}|m\rangle P_m(N-1),
\label{eq:energy2}
\earr
where we used $g(p)|n\rangle=g(p_n)|n\rangle$. We are mostly
interested in the evolution of the quantities $p$ and $p^2$
(momentum and kinetic energy). By the Baker-Hausdorff lemma
\andy{eq:baklemma}
\beq
U_{\rm kick}^\dagger g(p) U_{\rm kick}=g(p)+i\frac{\lambda}{\hbar}[V,g(p)]+\frac{1}{2!}\left(
\frac{i\lambda}{\hbar}\right)^2[V,[V,g(p)]]+...,
\label{eq:baklemma}
\eeq
we obtain the exact expressions
\andy{eq:dhzero,dhzeroo}
\barr
U_{\rm kick}^\dagger p U_{\rm kick} &=& p+i\frac{\lambda}{\hbar}[V,p],
\label{eq:dhzero} \\
U_{\rm kick}^\dagger p^2 U_{\rm kick} &=& p^2+i\frac{\lambda}{\hbar}[V,p^2]+
\lambda^2\left(V'\right)^2,
\label{eq:dhzeroo}
\earr
where prime denotes derivative. We observe, incidentally, that in
general, for  polynomial $g(p)$, the highest order of $\lambda$
appearing in (\ref{eq:baklemma}) is the degree of the polynomium.

Substituting (\ref{eq:dhzero}) and (\ref{eq:dhzeroo}) in
(\ref{eq:energy2}) and then iterating on the number of kicks we
obtain
\andy{eq:dhzero0,1}
\barr
& & \mean{p}_N =\mean{p}_{N-1}=\mean{p}_{0},\label{eq:dhzero0} \\
&
 & \mean{p^2}_N =\mean{p^2}_{N-1}+\lambda^2 \mean{f^2}
=\mean{p^2}_{0} +\lambda^2 \mean{f^2}N,
\label{eq:dhzero1}
\earr
where $f=-V'(x)$ is the force and
\andy{meanforce}
\beq
\mean{f^2}= {\rm Tr}\left(f^2\rho_{NT+\tau}\right)=
\sum_n\langle n|f^2|n\rangle P_n(N)=\frac{1}{2\pi}\int_{-\pi}^\pi dx\;f^2(x)
\label{eq:meanforce}
\eeq
is a constant that does not depend on $N$: Indeed $\langle
n|f^2|n\rangle$ is independent of the state $|n\rangle$ [see
(\ref{eq:eigenfun})] and $\sum P_n=1$. In particular, the kynetic
energy $K=p^2/2m$ grows at a constant rate:
$\mean{K}_{N}=\mean{K}_{0}+\lambda^2\mean{f^2} N/2m$. By using
(\ref{eq:dhzero0})-(\ref{eq:dhzero1}) we obtain the friction ($F$)
and the diffusion ($D$) coefficients
\andy{drift,diffusion}
\barr
& & F=\frac{\mean{p}_N-\mean{p}_0}{NT}=0,
\label{eq:drift} \\
& & D=\frac{\mean{\Delta p^2}_N-\mean{\Delta p^2}_0}{NT}
=\frac{\lambda^2 \mean{f^2}}{T},
\label{eq:diffusion}
\earr
where $\mean{\Delta p^2}_N=\mean{p^2}_N-\mean{p}_N^2$. We stress
that the above results are {\em exact}: their derivation involves
no approximation. This shows that Hamiltonian systems of the type
(\ref{eq:hamilt}) (radial twisting maps), in the quantum case, if
``measured" after every kick, have a constant diffusion rate in
momentum with no friction, for {\em any} perturbation $V=V(x)$. In
particular, the seminal kicked-rotator model $H_0=p^2/2I, V=\cos x$
has the diffusion coefficient
\andy{eq:krdiff}
\beq
D= \frac{\lambda^2}{2T},
\label{eq:krdiff}
\eeq
which is nothing but the result obtained in the classical case
\cite{Chirikov,KaulGontis}.

The above results are somewhat puzzling, essentially because one
finds that in the quantum case, when repeated measurements of
momentum (action variable) are performed on the system, a chaotic
behavior is obtained for {\em every} value of $\lambda$ and for
{\em any} potential $V(x)$. On the other hand, in the classical
case, diffusion occurs only for some $V(x)$, when $\lambda$ exceeds
some critical value $\lambda_{\rm crit}$. (For instance, the kicked
rotator  displays diffusion for $\lambda \geq \lambda_{\rm
crit}\simeq 0.972$ \cite{Chirikov,Lichten}.) It appears, therefore,
that quantum measurements not only yield a chaotic behavior in a
quantum context, they even produce chaos when the classical motion
is regular. In order to bring to light the causes of this peculiar
situation, one has to look at the classical case. The classical map
for the Hamiltonian (\ref{eq:hamilt}) reads
\andy{eq:classmap}
\barr
x_{N} &=& x_{N-1}+H'_0(p_{N-1})T, \nonumber \\
p_{N} &=& p_{N-1} -\lambda V'(x_{N}).
\label{eq:classmap}
\earr
A quantum measurement of $p$ yields an exact determination of
momentum $p$ and, as a consequence, makes position $x$ completely
undetermined (uncertainty principle). This situation has no
classical analog: it is inherently quantal. However, the classical
``map" that best mymics this physical picture is obtained by
assuming that position $x_N$ at time $\tau$ after each kick (i.e.\
when the quantum counterpart undergoes a measurement) behaves like
a random variable $\xi_N$ uniformly distributed over $[-\pi,\pi]$:
\andy{eq:classmeas}
\barr
x_{N}&=&\xi_{N}, \nonumber \\ p_{N}&=&p_{N-1}-\lambda V'(x_{N}).
\label{eq:classmeas}
\earr
Introducing the ensemble average $\langle\langle
\cdots\rangle\rangle$ over the stochastic process
(i.e.\ over the set of independent random variables $\{\xi_k
\}_{k\leq N}$),
we obtain
\andy{eq:meanp}
\beq
\langle\langle p_N \rangle\rangle =
\langle\langle p_{N-1} \rangle\rangle
-\lambda \langle V'(\xi_{N}) \rangle,
\label{eq:meanp}
\eeq
where
\andy{eq:meandef}
\beq
\langle g(\xi) \rangle \equiv
\frac{1}{2\pi}\int_{-\pi}^\pi g(\xi) d\xi
\label{eq:meandef}
\eeq
is the average over the single random variable $\xi$ [this
coincides with the quantum average: see for instance the last term
of (\ref{eq:meanforce})]. The average of $V'(\xi_N)$ in
(\ref{eq:meanp}) vanishes due to the periodic boundary
conditions on $V$, so that
\andy{eq:dhzero0cl}
\beq
\langle\langle p_N \rangle\rangle =
\langle\langle p_{N-1} \rangle\rangle,
\label{eq:dhzero0cl}
\eeq
which is the same as Eq.~(\ref{eq:dhzero0}). Moreover, using
(\ref{eq:classmeas}) and (\ref{eq:dhzero0cl}) we get
\andy{eq:meanp2}
\beq
\langle\langle \Delta p_N^2\rangle\rangle
=\langle\langle p_N^2\rangle\rangle-\langle\langle p_N\rangle\rangle^2
=\langle\langle\Delta p_{N-1}^2\rangle\rangle
+\lambda^2\langle V'(\xi_{N})^2\rangle-2\lambda\langle\langle
p_{N-1}\rangle\rangle\langle V'(\xi_{N})\rangle.
\label{eq:meanp2}
\eeq
In writing (\ref{eq:meanp2}), the average of $V'(\xi_N) p_{N-1}$
has been factorized because $p_{N-1}$ depends only on $\{\xi_k
\}_{k\leq N-1}$, as can be evinced from (\ref{eq:classmeas}).
Using again the periodic boundary condition on $V$, one finally
gets
\andy{eq:mean22}
\beq
\langle\langle \Delta p_N^2\rangle\rangle =
\langle\langle\Delta p_{N-1}^2\rangle\rangle
+\lambda^2 \langle f^2\rangle
\label{eq:mean22}
\eeq
and the momentum diffuses at the rate (\ref{eq:diffusion}), as in
the quantum case with measurements. We obtain in this case a
diffusion taking place in the whole phase space, without effects
due to the presence of adiabatic islands.

It is interesting to compare the different cases analyzed: (A) a
classical system, under the action of a suitable kicked
perturbation, displays a diffusive behavior if the coupling
constant exceeds a certain threshold (KAM theorem); (B) on the
other hand, in its quantum counterpart, this diffusion is always
suppressed. (C) The introduction of measurements between kicks
encompasses this limitation, yielding diffusion in the quantum
case. More so, diffusion takes place for any potential and all
values of the coupling constant (namely, even when the classical
motion is regular). (D) The same behavior is displayed by a
``randomized classical map," in the sense explained above. These
conclusions are sketched in Table~1.
\begin{center}
{\small {\bf Table 1}: Classical vs quantum diffusion}  \\  \quad \\
\begin{tabular}{|c|c|c|}
  \hline\hline
A &classical         & diffusion for $\lambda > \lambda_{\rm crit}$ \\
  \hline
B  &quantum           & no diffusion  \\
  \hline
C  &quantum + measurements & diffusion $\forall \lambda$ \\
  \hline
D  &classical + random     & diffusion $\forall \lambda$ \\
  \hline\hline
\end{tabular}
\end{center}

\setcounter{equation}{0}
\section{Semiclassical limit}
 \label{sec-semicl}
 \andy{semicl}

As we have seen, the effect of quantum measurements is basically
equivalent to a complete randomization of the classical angle
variable $x$, at least if one's attention is limited to the
calculation of the diffusion coefficient in the chaotic regime. One
might therefore naively think that the randomized classical map
(\ref{eq:classmeas}) and the quantum map with measurements
(\ref{eq:master}), (\ref{eq:dhzero0})-(\ref{eq:diffusion}) are
identical. This expectation would be wrong: there are in fact
corrections in $\hbar$. It is indeed straightforward, using
Eqs.~(\ref{eq:energy2})-(\ref{eq:baklemma}), to obtain in the
quantum case
\barr
\langle p^3\rangle_N &=& \langle p^3\rangle_{N-1}+3\lambda^2
\langle f^2\rangle\langle p\rangle_{N-1}+\lambda^3\langle f^3\rangle,
\nonumber\\
\langle p^4\rangle_N &=& \langle p^4\rangle_{N-1}+6\lambda^2
\langle f^2\rangle\langle p^2\rangle_{N-1}
+4\lambda^3\langle f^3\rangle\langle p\rangle_{N-1}
+\lambda^4\langle f^4\rangle +\lambda^2\hbar^2\langle
(f')^2\rangle. \nonumber \\
\earr
On the other hand, using (\ref{eq:classmeas}) and the periodic
boundary conditions, one gets for the randomized classical map
\barr
\langle\langle p_N^3\rangle\rangle
&=& \langle\langle p_{N-1}^3\rangle\rangle+3\lambda^2
\langle f^2\rangle\langle\langle p_{N-1}\rangle\rangle
+\lambda^3\langle f^3\rangle,
\nonumber\\
\langle\langle p_N^4\rangle\rangle
&=& \langle\langle p_{N-1}^4\rangle\rangle+6\lambda^2
\langle f^2\rangle\langle\langle p_{N-1}^2\rangle\rangle
+4\lambda^3\langle f^3\rangle\langle\langle p_{N-1}\rangle\rangle
+\lambda^4\langle f^4\rangle.
\earr
Hence the two maps have equal moments up to third order, while the
fourth moment displays a difference of order $O(\hbar^2)$:
\andy{fourth}
\beq
\langle p^4 \rangle_N - \langle p^4 \rangle_{N-1}
=\langle\langle p_N^4\rangle\rangle
-\langle\langle p_{N-1}^4\rangle\rangle
+\lambda^2\hbar^2\langle (f')^2\rangle.
\label{eq:fourth}
\eeq

In order to understand better the similarities and differences
between the two maps, as well as the quantum mechanical
corrections, we focus our attention on the particular case of the
kicked rotator $H_0=p^2/2$, $V(x)=\cos x$, which gives rise to the
so-called standard map
\andy{eq:standmap}
\barr
x_{N} &=& x_{N-1}+p_{N-1}T, \nonumber \\
p_{N} &=& p_{N-1} +\lambda \sin x_{N}.
\label{eq:standmap}
\earr
The conditional probability density $W_{\rm cl}$ that an initial
state $(p',x')$  evolves after one step into the final state
$(p,x)$ is, from (\ref{eq:standmap}),
\barr
W_{\rm cl}(p,x|p',x')&=&\delta(p-p'-\lambda\sin
x)\;\delta(x-x'-p'T)\nonumber\\
&=&\delta(p-p'-\lambda\sin[x'+p'T])\;\delta(x-x'-p'T).
\earr
This is a completely deterministic evolution. On the other hand, if
one randomizes the standard map, as in (\ref{eq:classmeas}),
\andy{eq:randstandmap}
\barr
x_{N} &=& \xi_{N}, \nonumber \\
p_{N} &=& p_{N-1} +\lambda \sin x_{N},
\label{eq:randstandmap}
\earr
the conditional probability density becomes
\andy{cltransprobtot}
\barr
W_{\rm cl}(p,x|p',x')=W_{\rm
cl}(p,x|p')=P(x)\;\delta(p-p'-\lambda\sin x)
=\frac{1}{2\pi}\delta(p-p'-\lambda\sin x)
\label{eq:cltransprobtot}
\earr
and is independent of the initial position $x'$. It is therefore
possible to describe the dynamics by considering only the momentum
distribution
\andy{cltransprob}
\barr
W_{\rm cl}(p|p') &=&\frac{1}{2\pi}\int_{-\pi}^\pi
dx\;\delta(p-p'-\lambda\sin x)
=\frac{1}{\lambda\pi}\int_{-1}^{+1}\frac{dy}{\sqrt{1-y^2}}\;
\delta\left(y-\frac{p-p'}{\lambda}\right)
\nonumber\\
&=&\frac{1}{\pi}\frac{1}{\sqrt{\lambda^2-(p-p')^2}}\;\theta(\lambda-|p-p'|).
\label{eq:cltransprob}
\earr
Notice that $W_{\rm cl}(p|p')$ is a function of the momentum
transfer $|\Delta p|=|p-p'|$ and vanishes for $|\Delta p|>\lambda$.

Consider now the kicked quantum rotator with measurements. From
Eq.\ (\ref{eq:transprob}), the transition probability reads
\andy{qutransprobdef}
\beq
W_{\rm q}(p=\hbar n|p'=\hbar n')=\frac{1}{\hbar}
W_{nn'}=\frac{1}{\hbar}\left|\langle n|e^{-i\lambda \cos x
/\hbar}|n'\rangle\right|^2
\label{eq:qutransprobdef}
\eeq
and by using the definition (\ref{eq:eigenfun}) one obtains
\andy{transbessel}
\barr
\langle n|e^{-i\lambda \cos x/\hbar}|n'\rangle&=&
\int_{-\pi}^{\pi} dx \langle n|x\rangle
e^{-i\lambda \cos x/\hbar} \langle x|n'\rangle
\nonumber\\
&=&\frac{1}{2\pi}\int_{-\pi}^{\pi} dx \;e^{-i(n-n')x}
e^{-i\lambda \cos x/\hbar}=i^{n-n'}
J_{n-n'}\left(\frac{\lambda}{\hbar}\right),
\label{eq:transbessel}
\earr
where $J_{m}(z)$ is the Bessel function of order $m$. Therefore, in
the quantum case, from (\ref{eq:qutransprobdef}) and
(\ref{eq:transbessel}), we can write
\andy{qutransprob}
\beq
W_{\rm q}(p=\hbar n|p'=\hbar n')=\frac{1}{\hbar}
J_{\nu}\left(\frac{\lambda}{\hbar}\right)^2 \qquad (\Delta
p=p-p'=\hbar \nu; \;\; \nu \equiv n-n').
\label{eq:qutransprob}
\eeq
There are two important differences between the classical case
(\ref{eq:cltransprob}) and its quantum counterpart
(\ref{eq:qutransprob}): i) the quantum mechanical transition
probability $W_{\rm q}$ admits only quantized values of momentum
$\hbar n$, while the classical one $W_{\rm cl}$ is defined on the
real line; ii) momentum can change by any value in the quantum case
(notice however that this occurs with very small probability for
$|\Delta p|=\hbar |\nu|\gg\lambda$ \cite{Chirikov}), while in the
classical case this change is strictly constrained by $|\Delta
p|\leq\lambda$. These features have an interesting physical
meaning: see Figure 1. The transition probability of classical
momentum appears as an ``average" of its quantum counterpart, which
explains the strong analogy discussed in Section 2. At the same
time, the quantum mechanical transition probability has a small
nonvanishing tail for $|\Delta p|=\hbar |\nu|>\lambda$: this is at
the origin of the difference (\ref{eq:fourth}).
\begin{figure}[t]
\epsfig{file=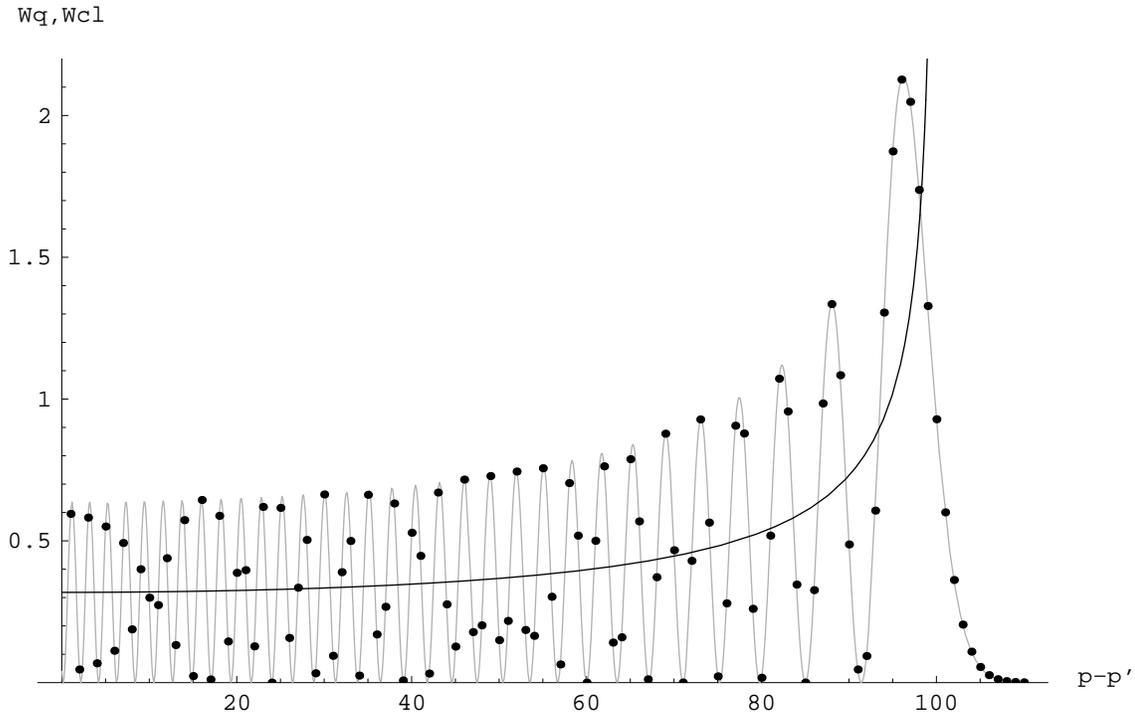, width=\textwidth}
\caption{Momentum transition probabilities for the kicked rotator
($\lambda = 100\hbar$ and the momentum transfer $p-p'$ is expressed
in units $\hbar$). The thick line is the classical expression
(\ref{eq:cltransprob}): it diverges for $p-p'=\lambda$ and vanishes
for $p-p' > \lambda$. The quantum mechanical transition probability
(\ref{eq:qutransprob}) is defined only for integer values of $p-p'$
(dots). The interpolating line (obtained by treating the order of
the Bessel function as a continuos variable) oscillates around its
classical counterpart and is nonvanishing (although very small)
outside the classical range, i.e.\ for $p-p' > \lambda$. }
\label{fig:bessel}
\end{figure}

Finally, let us show how one recovers the transition probability
$W_{\rm cl}$ starting from $W_{\rm q}$, in the semiclassical limit.
We look at the limit $\hbar\to0$, while keeping $\Delta p=\hbar\nu$
finite:
\andy{limit}
\beq
\hbar\to0,\quad\nu\to\infty\quad\mbox{with}\quad\Delta p=\hbar\nu=\mbox{const}.
\label{eq:limit}
\eeq
In this limit, the argument and the order of the Bessel function in
(\ref{eq:qutransprob}) are infinities of the same order. For
$|\Delta p|\leq\lambda$, setting $\Delta p/\lambda
\equiv\cos\beta$, one gets
\beq
\frac{\lambda}{\hbar}=\frac{\lambda}{\Delta p}\frac{\Delta p}{\hbar}
=\nu \sec\beta.
\eeq
Hence, by using the asymptotic limit of the Bessel function \cite{tables}
\andy{asympt}
\beq
J_{\nu}(\nu\sec\beta)
\stackrel{\nu\; {\rm large}}{\sim}
\sqrt{\frac{2}{\nu\pi\tan\beta}}
\left[\cos\left(\nu\tan\beta-\nu\beta-\frac{\pi}{4}\right)
+{\rm O}(\nu^{-1})\right],
\label{asympt}
\eeq
Eq.~(\ref{eq:qutransprob}) becomes, in the limit (\ref{eq:limit}),
\andy{semicltransprob}
\barr
& &W_{\rm q}(p|p')=\frac{1}{\hbar} J_{\frac{\Delta
p}{\hbar}}\left(\frac{\lambda}{\hbar}\right)^2=\frac{1}{\hbar}
J_{\nu}\left(\nu\sec\beta\right)^2\nonumber\\ & &\sim
\frac{1}{\hbar}\frac{2}
{\frac{\Delta p}{\hbar}\pi\sqrt{\frac{\lambda^2}{\Delta p^2}-1}}
\left[
\cos^2\left(\frac{\Delta p}{\hbar}\sqrt{\frac{\lambda^2}{\Delta p^2}-1}
-\frac{\Delta p}{\hbar}\arccos\frac{\Delta p}{\lambda}-\frac{\pi}{4}\right)
+{\rm O}\left(\frac{\hbar}{\Delta p}\right)\right]
\nonumber\\ & &\sim
W_{\rm cl}(p|p')\left[1+
\sin\left(\frac{2\sqrt{\lambda^2-\Delta p^2}}{\hbar}
-\frac{2\Delta p}{\hbar}\arccos\frac{\Delta p}{\lambda}\right)
+{\rm O}\left(\frac{\hbar}{\Delta p}\right)\right],
\nonumber \\
& & \hspace{8cm} (|\Delta p|\leq\lambda)
\label{eq:semicltransprob}
\earr
that, due to Riemann-Lebesgue lemma, tends to $W_{\rm cl}$ in the
sense of distributions.

On the other hand, for $|\Delta p|>\lambda$, setting $\Delta
p/\lambda \equiv \cosh\alpha$ and using the asymptotic formula \cite{tables}
\andy{asympt1}
\beq
J_{\nu}\left(\frac{\nu}{\cosh\alpha}\right)\stackrel{\nu\;
{\rm large}}{\sim}
\frac{\exp(\nu\tanh\alpha-\nu\alpha)}{\sqrt{2\nu\pi\tan\beta}}
\left[1+{\rm O}(\nu^{-1})\right],
\label{asympt1}
\eeq
we get
\andy{semicltransprob1}
\barr
& &W_{\rm q}(p|p')\nonumber\\ & &\sim
\frac{1}{2\pi\sqrt{\Delta p^2-\lambda^2}}
\exp\left\{-\frac{2\Delta p}{\hbar}
\left[\arccos\frac{\Delta p}{\lambda}
-\sqrt{1-\left(\frac{\lambda}{\Delta p}\right)^2}\right]
\right\}\left[1+{\rm O}\left(\frac{\hbar}{\Delta p}\right)\right],
\nonumber \\
& & \hspace{8cm} (|\Delta p|>\lambda)
\label{eq:semicltransprob1}
\earr
which vanishes exponentially (remember that $\tanh\alpha<\alpha$).
Equations (\ref{eq:semicltransprob}) and
(\ref{eq:semicltransprob1}) corroborate the results of Section 2
and enable us to conclude that the ``randomized" classical kicked
rotator is just the semiclassical limit of the ``measured" quantum
kicked rotator.

\setcounter{equation}{0}
\section{Concluding remarks }
 \label{sec-conclrem}
 \andy{conclrem}
The conclusion drawn in the previous section for the kicked rotator
can be generalized to an arbitrary radial twisting map. The
calculation and the techniques utilized are more involved and will
be presented elsewhere. There are also a number of related problems
that deserve attention and a careful investigation. Among these, we
just mention the case of imperfect quantum measurements, yielding a
partial loss of quantum mechanical coherence, the relation to
disordered systems, Anderson localization \cite{Flores} and quantum
Zeno effect \cite{QZE} and finally the extension to a different
class of Hamiltonians \cite{Casati}.

\end{document}